\documentclass[journal,12pt,draftclsnofoot,onecolumn,twoside]{IEEEtran}%

\usepackage{cite}

\ifCLASSINFOpdf
  \usepackage[pdftex]{graphicx}
\else
\fi
\graphicspath{{figures/}}

\usepackage[cmex10]{amsmath}
\usepackage{amssymb}
\usepackage{mathtools}

\usepackage{algorithm}
\usepackage{algpseudocode}
\usepackage[small]{caption}
\usepackage[font=footnotesize]{subfig}
\usepackage{epstopdf}					
\usepackage{amsthm}

\newtheorem{lemma}{Lemma}
\theoremstyle{plain}

\usepackage{bm}

\usepackage{xcolor}

\usepackage{hyperref}

\begin{document}
%
\title{\LARGE{A Very Low Complexity Successive Symbol-by-Symbol Sequence Estimator for  Faster-than-Nyquist Signaling}}
%
%
%

\author{{\normalsize{Ebrahim Bedeer, 
        Mohamed Hossam Ahmed, 
        and~Halim Yanikomeroglu 
}}
\thanks{This work is supported in part by DragonWave Inc., in part by the Mathematics of Information Technology and Complex Systems (MITACS) Canada, and in part by Natural Science and Engineering Research Council of Canada (NSERC) through Discovery program.}
\thanks{E. Bedeer  and H. Yanikomeroglu are with the System and Computer Engineering Department, Carleton University, Ottawa, ON K1S 5B6, Canada (e-mails: \{e.bedeer, halim.yanikomeroglu\}@sce.carleton.ca).}
\thanks{M. H. Ahmed is with the Faculty of 	Engineering and Applied Science, Memorial University, St. John’s, NL, A1B 3X5, Canada (e-mail: mhahmed@mun.ca).}
}

\maketitle

\begin{abstract}
In this paper, we investigate the sequence estimation problem of binary and quadrature phase shift keying faster-than-Nyquist (FTN) signaling and propose two novel low-complexity sequence estimation techniques based on concepts of successive interference cancellation.
To the best of our knowledge, this is the first approach in the literature to detect FTN signaling on a symbol-by-symbol basis.
In particular, based on the structure of the self-interference inherited in FTN signaling, we first find the operating region boundary---defined by the root-raised cosine (rRC) pulse shape, its roll-off factor, and the time acceleration parameter of the FTN signaling---where perfect estimation of the transmit data symbols on a symbol-by-symbol basis is guaranteed, assuming noise-free transmission. 
For noisy transmission, we then propose a novel low-complexity technique that works within the operating region  and is capable of estimating the transmit data symbols on a symbol-by-symbol basis. 
To reduce the error propagation of the proposed successive symbol-by-symbol sequence estimator (SSSSE), we propose a successive symbol-by-symbol with go-back-$K$ sequence estimator (SSSgb$K$SE)  that goes back to re-estimate up to $K$ symbols,  and subsequently improves the estimation accuracy of the current data symbol.
Simulation results show that the proposed sequence estimation techniques perform well for low intersymbol interference (ISI) scenarios and can significantly increase the data rate and spectral efficiency. Additionally, results reveal that choosing the value of $K$ as low as $2$ or $3$ data symbols is sufficient to  significantly improve the bit-error-rate  performance. 
Results also show that the performance of the proposed SSSgb$K$SE, with $K = 1$ or $2$, surpasses the performance of the lowest complexity equalizers reported in the literature, with reduced computational complexity.

\end{abstract} 



\section{Introduction} \label{sec:into}

There is a need to design better spectral efficient digital communication systems, as data rate requirements are conservatively doubling each year. The term Faster-than-Nyquist (FTN) signaling was coined in late 60s and early 70s \cite{saltzberg1968intersymbol, lucky1970decision, salz1973optimum}  to refer to digital transmission of pulses beyond the Nyquist limit. The pioneering work of J. E. Mazo in 1975 \cite{mazo1975faster} was the first to prove that FTN signaling does not affect the minimum distance of binary sinc pules when transmitted at rate $\frac{1}{\tau}$, $\tau \in [0.802, 1]$, higher than the Nyquist signaling; this is called the Mazo limit. In other words, Mazo proved that almost $25\%$ more bits, compared to the Nyquist signaling, can be transmitted in the same bandwidth and at the same signal-to-noise ratio (SNR) without degrading the bit error rate (BER), assuming perfect processing at the receiver. 



Despite the doubts raised by Foschini on the benefits of FTN signaling in \cite{foschini1984contrasting}, its potential to improve the transmission rates was revealed in \cite{rusek2006cth04, rusek2009constrained}. Although, implementations of  FTN signaling in practical communication systems pose several challenges mainly due to the high complexity involved to remove the self-interference. 
In \cite{liveris2003exploiting}, the binary FTN signaling was viewed as  a convolutionally encoded transmission and a Viterbi algorithm (VA) was used for detection. To reduce the complexity of the  FTN signaling detection problem in \cite{liveris2003exploiting}, truncated VA  \cite{prlja2008receivers} and reduced states Bahl-Cocke-Jelinek-Raviv (BCJR) algorithm \cite{anderson2009new} are investigated; yet,  the works in \cite{prlja2008receivers, anderson2009new} are still complex and more effective for strong ISI scenarios. 
For low ISI scenarios,  
a frequency domain equalizer (FDE)  has been proposed in \cite{sugiura2013frequency} to detect  FTN signaling with reasonable complexity. However, due to the insertion of a guard interval, e.g. cyclic prefix, at the transmitter, the effective transmission rate is reduced depending on the relative length of the cyclic prefix and the transmitted data block. 
For instance, for a data block transmission of 1000 symbols and a cyclic prefix of 10 symbols, the effective transmission rate is reduced by $1\%$.
The work in \cite{sugiura2013frequency} was extended in \cite{sugiura2015frequency} to provide soft decisions about the data symbols using FDE-aided three-stage concatenated turbo decoder. The works in \cite{sugiura2013frequency, sugiura2015frequency} were extended to produce soft-decision of the estimated data symbols while considering the correlated noise samples after the receiver matched filter in   \cite{ishihara2016frequency}. 
In \cite{dinis2015hybrid}, the authors proposed an iterative block decision feedback frequency domain equalizer in addition to a hybrid automatic repeat request to detect FTN signaling with reduced computational complexity. 

Other important aspects of FTN signaling have been recently studied as well. For instance, the authors in \cite{lucciardi2016trade} studied the tradeoff between increasing the spectral efficiency (SE) of FTN signaling, as a result of the accelerated pulses' transmission in time, and increasing the FTN signaling peak to average power ratio. In \cite{le2015faster}, the effectiveness of multicarrier FTN signaling is investigated for coexistence scenarios. In particular, it was shown that multicarrier FTN signaling can compensate for the loss in SE due to using guard bands between different systems.

To the best of the authors' knowledge, this is the first approach in the literature to detect FTN signaling on a symbol-by-symbol basis. This is in contrast to all the previous sequence estimation techniques reported in the literature that estimate the transmit data symbols based on the reception of a block of length $N$ \cite{liveris2003exploiting, prlja2008receivers, anderson2009new, sugiura2013frequency, sugiura2015frequency, ishihara2016frequency, dinis2015hybrid}.
The main contributions of this paper are summarized as follows:
\begin{itemize}
	\item We exploit the structure of the self-interference inherited in binary and quadrature phase shift keying (BPSK and QPSK) FTN signaling to find the operating region boundary where perfect data symbols estimation on a symbol-by-symbol basis is guaranteed, assuming noise-free transmission. The operating region boundary is defined by the root-raised cosine (rRC) pulse shape, its roll-off factor $\beta$,  and the time acceleration parameter $\tau$ of the FTN signaling.
	\item For noisy transmission, we propose a successive symbol-by-symbol sequence estimator (SSSSE)  that works within the operating region and is capable of estimating the transmit data symbols in a low-complexity manner.
	\item We additionally propose a successive symbol-by-symbol with go-back-$K$ sequence estimator (SSSgb$K$SE)  to reduce the error propagation effect of the proposed SSSSE. The proposed SSSgb$K$SE  can go back to re-estimate up to  $K$ data symbols (based on the knowledge of the current data symbol and its previous $K - 1$ data symbols), and subsequently improves the estimation accuracy of the current data symbol (based on the re-estimation of the previous $K$ data symbols).  
	\item Simulation results show the effectiveness of the proposed SSSSE and SSSgb$K$SE techniques for low ISI scenarios to considerably increase the data rate and SE. Additionally, results reveal that for low ISI scenarios choosing the value of $K$ as low as $2$ or $3$ data symbols is sufficient to  significantly improve the BER performance.  Results also show that the proposed SSSgb$K$SE, with $K = 1$ or $2$, outperforms the lowest complexity equalization techniques reported in the open literature, with reduced computational complexity.
\end{itemize}

The remainder of this paper is organized as follows. Section \ref{sec:model} presents the system model of the FTN signaling. The proposed SSSSE is discussed in Section \ref{sec:SSD}, while the proposed SSSgb$K$SE is introduced in Section \ref{sec:SSSgb$K$SE}. Section \ref{sec:results} provides the performance results of our proposed sequence estimation techniques, and finally the paper is concluded in Section \ref{sec:conc} 
\section{FTN Signaling System Model} \label{sec:model} 

\begin{figure}[!t]
	\centering
	\includegraphics[width=1.00\textwidth]{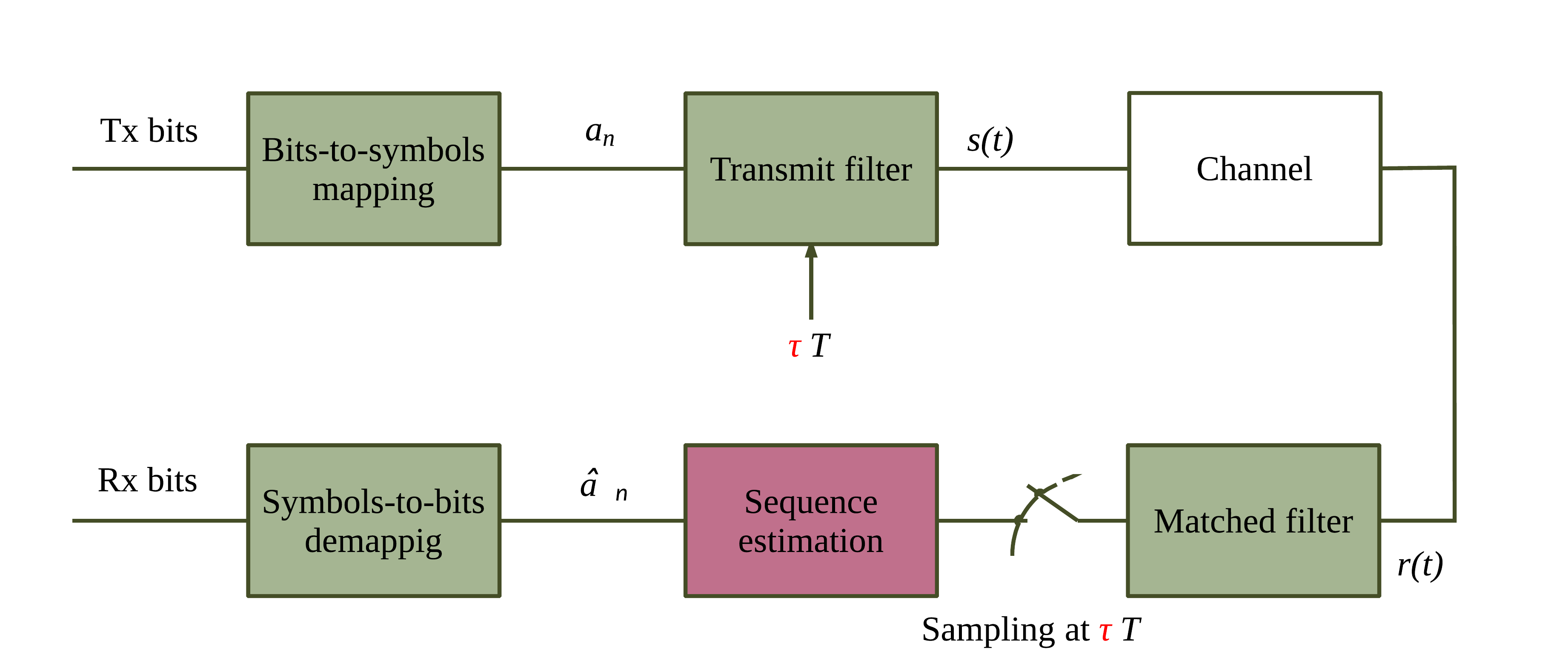}
	\caption{Block diagram of FTN signaling.}\label{fig:block_diagram}
\end{figure}

Fig. \ref{fig:block_diagram} shows a  block diagram of a communication system employing FTN signaling. Data bits to be transmitted are gray mapped\footnote{It is worthy to mention that there may exist other bits-to-symbol mapping schemes that result in better performance of the FTN signaling; however, investigating such schemes are out of the scope of this paper.} to data symbols through the bits-to-symbols mapping block. Data symbols are transmitted, through the rRC transmit filter block, faster than Nyquist signaling, i.e., every $\tau T$, where $0 < \tau \leq 1$ is the time packing/acceleration parameter and $T$ is the symbol duration. A possible receiver structure is shown in Fig. \ref{fig:block_diagram}, where the received signal is passed through a filter matched to the rRC transmit filter followed by a sampler. Since the transmission rate of the transmit pulses carrying the data symbols intentionally violate the Nyquist criterion, ISI occurs between the received samples. Accordingly, sequence estimation techniques are needed to remove the ISI and to estimate the transmitted data symbols. The estimated data symbols are finally gray demapped to the estimated received bits.

The transmitted signal $s(t)$ of the FTN signaling shown in Fig. \ref{fig:block_diagram} can be written in the form
\begin{eqnarray}
s(t) = \sqrt{E_s} \: \sum\nolimits_{n = 1}^{N} a_n \: p(t - n \tau T), \qquad 0 < \tau \leq 1,
\end{eqnarray} 
where $N$ is the total number of transmit data symbols, $a_n, \: n= 1, \hdots, N,$ is the independent and identically distributed  data symbols, $E_s$ is the data symbol  energy, $p(t)$ is a unit-energy pulse, i.e., $\int\nolimits_{-\infty}^{\infty} \vert p(t) \vert^2 dt = 1$,  and $1/(\tau T)$ is the signaling rate. 
The received FTN signal in case of additive white Gaussian noise (AWGN) channel is written as
\begin{eqnarray}
y(t) = s(t) + n(t), 
\end{eqnarray}
where $n(t)$ is a zero mean complex valued Gaussian random variable with variance $\sigma^2$. A possible receiver architecture for FTN signaling is to use a filter matched to $p(t)$; thus the received signal after the matched filter can be written as
\begin{eqnarray}
y(t) = \sqrt{E_s} \: \sum\nolimits_{n = 1}^{N} a_n g(t - n \tau T) + w(t),
\end{eqnarray}
where $g(t) = \int\nolimits p(x) p(x - t) dx$ and $w(t) = \int\nolimits n(x) p(x - t) dx$. 
Assuming perfect timing synchronization between the transmitter and the receiver, the received FTN signal $y(t)$ is sampled every $\tau T$ and  the $k$th received sample can be expressed as
\begin{eqnarray}\label{eq:ISI_symbols}
y_k &=& y(k \tau T) \nonumber \\
&=& \sqrt{E_s} \sum\nolimits_{n = 1}^{N} a_n g(k \tau T - n \tau T) + w(k \tau T) \nonumber \\
&=& \underbrace{\sqrt{E_s} \: a_k \: g(0)}_{\textup{desired symbol}} + \underbrace{\sqrt{E_s} \: \sum\nolimits_{n = 1, \: n \ne k}^{N} a_n \: g((k - n) \tau T)}_{\textup{ISI from adjacent symbols}} + w(k \tau T).
\end{eqnarray} 

The optimal detector of the FTN signaling in \eqref{eq:ISI_symbols} in terms of minimizing the block-error-rate  is the maximum likelihood sequence estimation; however, its NP-hard computational complexity is  prohibitive for practical implementations. In the following, we discuss and propose  very low complexity symbol-by-symbol sequence estimators for BPSK and QPSK FTN signaling.

\section{Successive Symbol-by-Symbol Sequence Estimation (SSSSE)} \label{sec:SSD}

As discussed earlier, all the previous FTN signaling sequence estimation techniques reported in the literature  estimate the transmit data symbols based on the reception of a block of length $N$ \cite{liveris2003exploiting, prlja2008receivers, anderson2009new, sugiura2013frequency, sugiura2015frequency, ishihara2016frequency, dinis2015hybrid}. In this section, we propose a novel SSSSE that  estimates the transmit data symbols  on a symbol-by-symbol basis. 

The key enabler behind the proposed SSSSE is an observation about an operation region of BPSK and QPSK FTN signaling, where perfect estimation of data symbols on a symbol-by-symbol basis is guaranteed for noise-free transmission. The boundary of this operation region is defined by the rRC pulse shape, its roll-off factor $\beta$, and the time acceleration parameter $\tau$. 
To explain the basic idea of the observation that lead to the proposed SSSSE, let us rewrite \eqref{eq:ISI_symbols} in a vector form for noise-free transmission as 
\begin{eqnarray}\label{eq:ISI_matrix}
\bm{y} &=& \bm{G} \: \bm{a}  \nonumber \\
\begin{bmatrix}
y_1\\ 
y_2\\ 
y_3\\ 
\vdots\\
y_k\\ 
\vdots\\ 
y_N\\
\end{bmatrix}&=& \begin{bmatrix}
G_{1,1} & G_{1,2}  & G_{1,3}  & \hdots  & G_{1,L} & 0 & 0 & 0 & 0 \\ 
G_{1,2} & G_{1,1}  & G_{1,2}  & G_{1,3} & \hdots  & G_{1,L} & 0 & 0 & 0  \\ 
G_{1,3} & G_{1,2} & G_{1,1}  & G_{1,2}  & G_{1,3} & \hdots  & G_{1,L} & 0 & 0   \\ 
\ddots & \ddots  & \ddots & \ddots  & \ddots  & \ddots & \ddots & \ddots &     \ddots\\ 
\hdots & \hdots  & G_{1,3} & G_{1,2}  & G_{1,1} & G_{1,2}  & G_{1,3} & \hdots &     \hdots\\ 
\ddots & \ddots & \ddots & \ddots & \ddots & \ddots & \ddots  & \ddots &     \ddots\\ 
0 & 0 & 0 & 0 & G_{1,L} & \hdots & G_{1,3} & G_{1,2} & G_{1,1}
\end{bmatrix}
\begin{bmatrix}
a_1\\ 
a_2\\ 
a_3\\ 
\vdots\\
a_k\\ 
\vdots\\
a_N\\
\end{bmatrix},
\end{eqnarray}
where $\bm{G}$ is the ISI matrix, $G_{n,n'} = g((n - n') \tau T)$ represents the ISI between data symbols $n$ and $n'$, and $L - 1$ is the length of the one-sided ISI.
As can be seen in \eqref{eq:ISI_matrix}, each received sample value, i.e., $y_k$, is affected by ISI  in the form of an accumulated weighted sum of up to $2 L - 2$ adjacent data symbols. This ISI depends on the rRC pulse shape, its roll-off factor $\beta$, and the time acceleration parameter $\tau$ of the FTN signaling. 
Following \eqref{eq:ISI_matrix},   the $k$th received sample is expressed as 
\begin{eqnarray}
y_{k} = \underbrace{G_{1, L} \: a_{k - L + 1} + \hdots + G_{1, 2} \: a_{k - 1}}_{\textup{ISI from previous $L - 1$ symbols}} + \underbrace{G_{1,1} \: a_{k}}_{\textup{Current symbol to be estimated}} + \underbrace{G_{1, 2} \: a_{k + 1} + \hdots + G_{1, L} \: a_{k + L - 1}}_{\textup{ISI from upcoming $L - 1$ symbols}}.
\end{eqnarray}
Hence, to detect the $k$th transmit symbol $a_k$ from the $k$th received sample $y_k$, we need to remove the interference from the previously detected $L - 1$ data symbols $a_{k - 1}, \hdots, a_{k - L + 1}$ and from the upcoming $L - 1$ (and yet undetected)  data symbols $a_{k + 1}, \hdots, a_{k +  L - 1}$. That said, the main challenge is how to handle the interference from the upcoming $L - 1$ data symbols, while still estimating the current data symbol $a_k$ based on a symbol-by-symbol basis. 
In the following, we explain how to handle the interference from the upcoming $L-1$ data symbols for the case of BPSK and QPSK FTN signaling. The same idea can be extended to higher constellations as well.

\subsection{Binary Phase Shift Keying FTN Signaling}

For BPSK FTN signaling, the perfect reconstruction condition is outlined in \textbf{Lemma} \ref{lemma:BPSK_condition} below.


\begin{figure}[!t]
	\centering
	\includegraphics[width=1.00\textwidth]{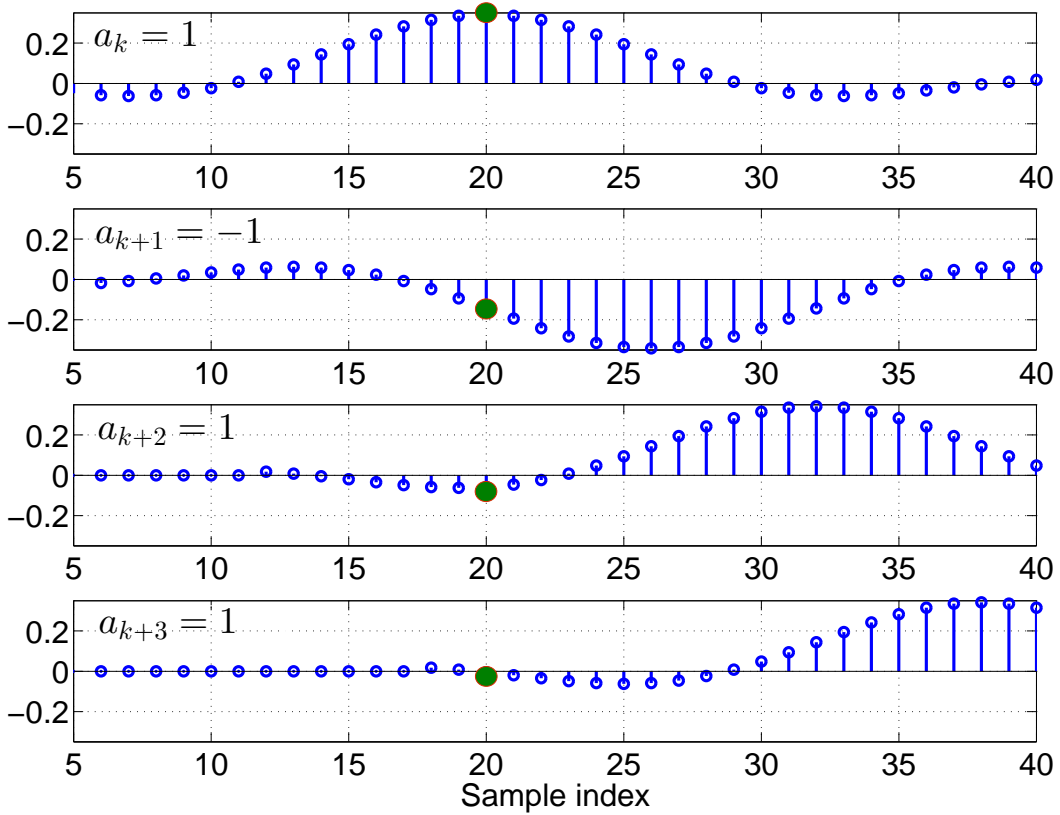}
	\caption{Explanation of the basic idea of the proposed SSSSE.}\label{fig:explain}
\end{figure}

\begin{lemma}\label{lemma:BPSK_condition}
Perfect estimation condition for BPSK FTN signaling for noise-free transmission.

Regardless the value of the current data symbol $a_k$, the upcoming $L - 1$ data symbols $a_{k + 1}, \hdots, a_{k +  L - 1}$, and the value of $L$, the following inequality holds for a certain range of $\tau$ and $\beta$:
\begin{eqnarray}\label{eq:inequality}
\vert G_{1,1}  \: a_k \vert > \vert G_{1,2} \: a_{k + 1} + \hdots + G_{1,L} \: a_{k +  L - 1} \vert.
\end{eqnarray}
\end{lemma}

\emph{Proof:} the range of $\tau$ and $\beta$ that satisfies the perfect estimation condition in \eqref{eq:inequality} can be found by a simple numerical search on the upcoming $L - 1$ data symbols $a_{ k + 1}, \hdots, a_{k +  L - 1}$ that will result in the worst ISI contribution to the current data symbol $a_k$ as follows.
The worst ISI scenario for the estimation of $a_k$ occurs when the upcoming $L -1$ data symbols $a_{k + 1}, ..., a_{k +  L - 1}$ are chosen such that  $G_{1, 2} \: a_{k + 1},  \hdots,  G_{1, L} \: a_{k +  L - 1}$ has an opposite sign to $G_{1, 1} \: a_k$. In other words, based on the signs of $G_{1,2}, \hdots, G_{1, L}$, the data symbols $a_{k + 1}, \hdots, a_{k +  L - 1}$ are  chosen such that the ISI to the $k$th received sample is maximized. This can be illustrated with  the help of Fig.~\ref{fig:explain}, where the $k$th transmit data symbol $a_k$ is affected by the interference from  the upcoming $L - 1 = 3$ transmit data symbols. Without loss of generality,  we assume that the $k$ transmit data symbol $a_k$ in Fig.~\ref{fig:explain}~(a) is $+1$. 
Given that the signs of $G_{1, 2}, G_{1, 3}$, and $G_{1, 4}$ are positive, negative, and negative respectively, then the worst interference affecting $a_k$ will occur when $a_{k + 1} = -1$, $a_{k + 2} = 1$, and $a_{k + 3} = 1$ such that $G_{1, 2} \: a_k$, $G_{1, 3} \: a_{k + 1}$, and $G_{1, 4} \: a_{k + 2}$ are all negative values (i.e., opposite to $a_k$), and hence, the interference to the $k$th data symbol, i.e., $G_{1, 2} \: a_k +  G_{1, 3} \: a_{k + 1} + G_{1, 4} \: a_{k + 2}$, is maximized. On the other hand, if the $k$th transmit data symbol $a_k$ is $- 1$, then we choose $a_{k + 1} =  1$, $a_{k + 2} = - 1$, and $a_{k + 3} = - 1$ such that $G_{1, 2} \: a_k$, $G_{1, 3} \: a_{k + 1}$, and $G_{1, 4} \: a_{k + 2}$ have all positive values (i.e., opposite to $a_k$), and hence, the interference to the $k$th data symbol, i.e., $G_{1, 2} \: a_k +  G_{1, 3} \: a_{k + 1} + G_{1, 4} \: a_{k + 2}$, is maximized. \hfill$\blacksquare$



\subsection{Quadrature Phase Shift Keying FTN Signaling}
Similar to the discussion of the BPSK FTN signaling, the perfect estimation condition for QPSK FTN signaling is outlined in \textbf{Lemma} \ref{lemma:QPSK_condition} below.

\begin{lemma}\label{lemma:QPSK_condition}
	Perfect estimation condition for QPSK FTN signaling for noise-free transmission.
	
	Regardless the value of the current data symbol $a_k$, the upcoming $L - 1$ data symbols $a_{k + 1}, \hdots, a_{k +  L - 1}$, and the value of $L$, the following inequality holds for a certain range of $\tau$ and $\beta$:
\begin{eqnarray}\label{eq:inequality_QPSK}
\vert G_{1,1}  \: \Re\{a_k\} \vert > \vert G_{1,2} \: \Re\{a_{k + 1}\} + \hdots + G_{1,L} \: \Re\{a_{k +  L - 1}\} \vert, \label{eq:inequality_QPSK} \\
\vert G_{1,1}  \: \Im\{a_k\} \vert > \vert G_{1,2} \: \Im\{a_{k + 1}\} + \hdots + G_{1,L} \: \Im\{a_{k +  L - 1}\} \vert, \label{eq:inequality_QPSK2}
\end{eqnarray}
where $\Re\{.\}$ and $\Im\{.\}$ are the real and imaginary parts of a complex number.
\end{lemma}

\emph{Proof:} \textbf{Lemma} \ref{lemma:QPSK_condition}  can be proved similar to \textbf{Lemma} \ref{lemma:BPSK_condition}; hence, the proof is omitted to avoid unnecessary repetitions. \hfill$\blacksquare$

\subsection{Finding the Operation Region Boundary}

To find the range of $\beta$ and $\tau$ such that the perfect estimation conditions in \textbf{Lemma} \ref{lemma:BPSK_condition} and \textbf{Lemma} \ref{lemma:QPSK_condition} hold, and hence, perfect estimation of data symbols on symbol-by-symbol basis is guaranteed for noise-free transmission, we perform the following offline search. For BPSK FTN signaling and for a certain value of $\beta$ and $\tau$ and the ISI samples between adjacent symbols, i.e. $G_{1, 1}, \hdots, G_{1, L}$, we select the upcoming $L - 1$  symbols $a_{k + 1}, \hdots, a_{k +  L - 1}$ according to the  the signs of $G_{1, 2}, \hdots, G_{1, L}$, respectively. 
For instance, for the case of $a_k = +1$, the upcoming $L - 1$ data symbols $a_{k + 1}, \hdots, a_{k +  L - 1}$ are selected to be of opposite sign to $G_{1, 2}, \hdots, G_{1, L}$, respectively. On the other hand, for the case of  $a_k = -1$, the upcoming $L - 1$ data symbols $a_{k + 1}, \hdots, a_{k +  L - 1}$ are selected to be of the same sign to $G_{1, 2}, \hdots, G_{1, L}$, respectively.
We note that $G_{1, 2} \: a_{k + 1} + \hdots + G_{1, L} \: a_{k + L - 1}$ represents the worst ISI that can affect the $k$th transmit data symbol $a_k$. Then, we gradually decrease the value of $\tau$ until \textbf{Lemma} \ref{lemma:BPSK_condition} and \textbf{Lemma} \ref{lemma:QPSK_condition} are violated. We follow similar approach for the case of QPSK FTN signaling to find the value of $\tau$. This is formally expressed as follows:


{\textit{Algorithm 1:} Finding the Operation Region Boundary}
\begin{enumerate}
	\item \textbf{Input:} rRC pulse shape and its roll-off factor $\beta$.
	\item Initialize the value of $\tau = 1$. 
	\item Calculate the values of $G_{1,1}, \hdots, G_{1, L}$.
	\item Select $a_{k + 1}, \hdots, a_{k +  L - 1}$ to have the same signs as $G_{1,1}, \hdots, G_{1, L}$, respectively.
	\item Decrease the value of $\tau$ until the perfect estimation conditions in \textbf{Lemma} \ref{lemma:BPSK_condition} and \textbf{Lemma} \ref{lemma:QPSK_condition} are violated.
	\item \textbf{Output:} Final value of $\tau$.
\end{enumerate}


Following {\textit{Algorithm 1}, the operation region and the SE bound, where  perfect data symbols estimation on a symbol-by-symbol basis is guaranteed for noise-free transmission, are summarized in Table \ref{table:region}. { It is worthy to emphasize that the operating region is found for the noise-free transmission scenario. It is expected that the proposed schemes working in a noisy transmission will give unsatisfactory performance if the operating parameters $\tau$ and $\beta$ are selected to be  at the edge of the operating region. This is as the noise can easily violate the perfect reconstruction conditions and move the proposed schemes operation outside the operating region.}
As expected, the operation region boundaries match for both BPSK and QPSK FTN signaling. 
For the reader's convenience, the operation region and the SE bound of QPSK FTN signaling is plotted in Fig. \ref{fig:SE_3D}. 


\begin{table}[]
	\centering
	\caption{Operating region boundary and the SE bound.}
	\label{table:region}
\begin{tabular}{c|c|c|c} \hline 
	\multicolumn{1}{c|}{$\beta$} & \multicolumn{1}{c|}{$\tau$} & \multicolumn{1}{c|}{\begin{tabular}[c]{@{}c@{}}SE bound (bits/sec/Hz)\\ BPSK FTN\end{tabular}} & \multicolumn{1}{c}{\begin{tabular}[c]{@{}c@{}}SE bound (bits/sec/Hz)\\ QPSK FTN\end{tabular}} \\ \hline \hline 
	0 & 0.68 & 1.47 & 2.94 \\
	0.1 & 0.63 & 1.44 & 2.89 \\
	0.2 & 0.59 & 1.41 & 2.82 \\
	0.3 & 0.49 & 1.57 & 3.14 \\
	0.4 & 0.47 & 1.52 & 3.03 \\
	0.5 & 0.45 & 1.48 & 2.96 \\
	0.6 & 0.43 & 1.45 & 2.90 \\
	0.7 & 0.41 & 1.43 & 2.87 \\
	0.8 & 0.39 & 1.42 & 2.85 \\
	0.9 & 0.37 & 1.42 & 2.85 \\
	1 & 0.35 & 1.43 & 2.86 \\ \hline
\end{tabular}
\end{table}

\begin{figure}[!t]
	\centering
	\includegraphics[width=1.00\textwidth]{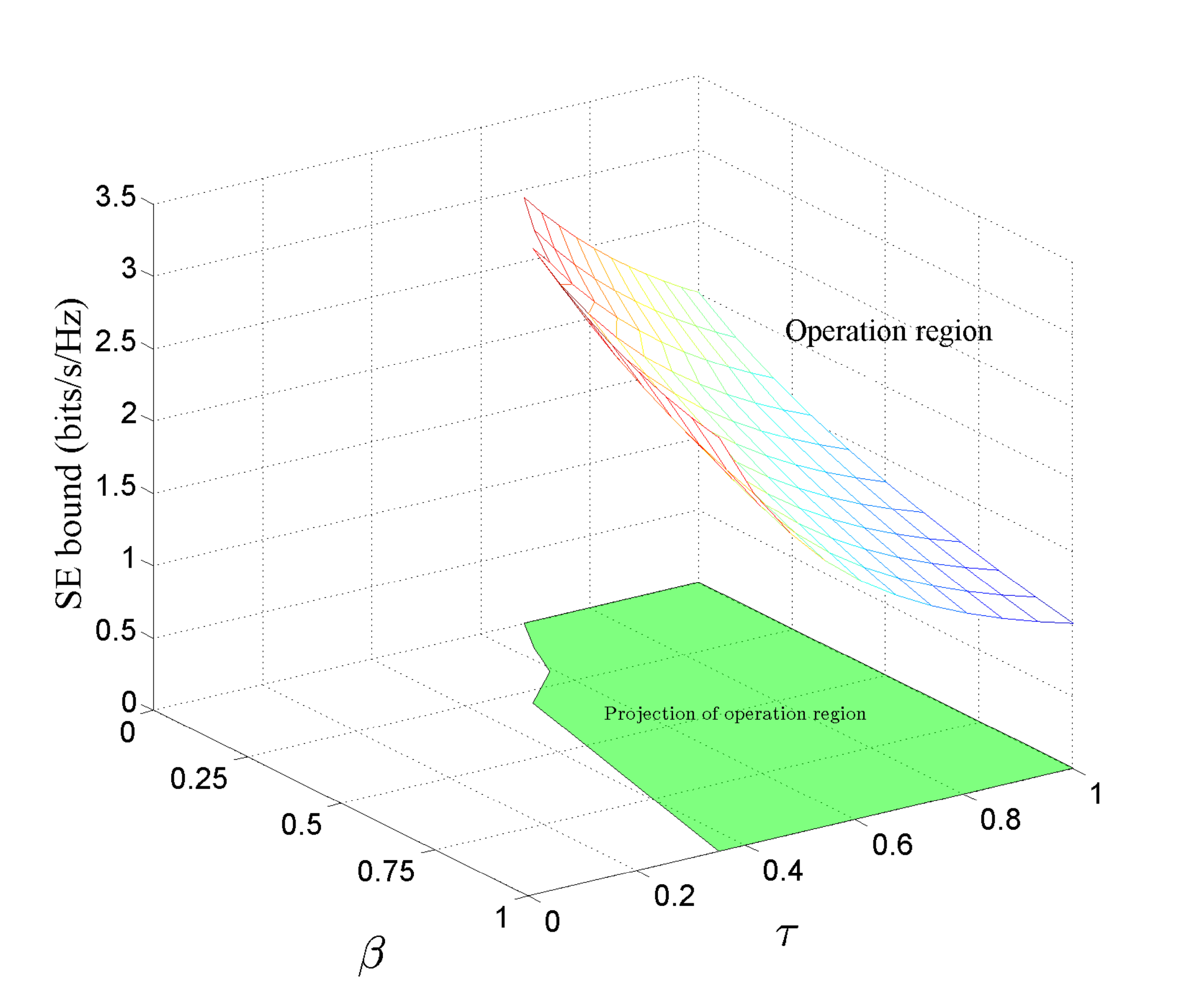}
	\caption{Operation region and SE bound of  \textbf{Lemma} \ref{lemma:QPSK_condition}, where perfect data symbols estimation on a symbol-by-symbol basis is guaranteed for noise-free transmission.}\label{fig:SE_3D}
\end{figure}	

Under these operating conditions, the $k$th data symbol $a_{k}$ can be estimated on a symbol-by-symbol basis as follows. First, the contribution from the previous $L - 1$ data symbols are subtracted from the $k$th received sample $y_k$. Then, the $k$ data symbol $a_k$ is estimated in the presence of the ISI from the upcoming $L - 1$ data symbols and thermal noise as 
\begin{eqnarray}\label{eq:estimation_rule}
\hat{a}_k = {\rm{quantize}} \left\{y_k - (G_{1, L} \: \hat{a}_{k - L + 1} + \hdots + G_{1, 2} \: \hat{a}_{k - 1})\right\},
\end{eqnarray}
where ${\rm{quantize}}\{x\}$ rounds  $x$ to the nearest BPSK/QPSK symbol\footnote{It is worthy to mention that the proposed schemes provide only hard decisions about the data symbols. One possible way to provide soft decisions about the data symbols is to approximate the ISI (from previous and upcoming symbols) as a zero-mean Gaussian process with a given variance according to the ISI term in \eqref{eq:ISI_symbols}. Then assume that the received samples are affected by zero-mean Gaussian process with variance equals to the sum of noise and ISI variances.}.

\subsection{Proposed SSSSE  and Complexity Analysis}
The proposed SSSSE  can  formally be expressed as follows:

{\textit{Algorithm 2:} Proposed SSSSE }
\begin{enumerate}
	\item \textbf{Input:} rRC pulses shape, its roll-off factor $\beta$, and the time acceleration parameter $\tau$.
	\item \textbf{If} $\beta$ and $\tau$ belong to the operation region shown in Table \ref{table:region} \textbf{then}
	\item \hspace{1cm} $\hat{a}_k = {\rm{quantize}} \left\{y_k - (G_{1, L} \: \hat{a}_{k - L + 1} + \hdots + G_{1, 2} \: \hat{a}_{k - 1})\right\}$.
	\item \textbf{End}
\end{enumerate}	


When compared to Nyquist signaling, the proposed SSSSE requires to subtract the effect of the ISI of the previous $L - 1$ data symbols, as can be seen in \eqref{eq:estimation_rule}. This translates to additional $L-2$ additions and $L - 1$ multiplications operations, when compared to the Nyquist signaling detection. 
\section{Successive Symbol-by-Symbol with go-back-$K$ Sequence Estimation (SSSgb$K$SE)} \label{sec:SSSgb$K$SE}

The proposed SSSSE suffers from the error propagation effect, i.e., an incorrectly estimated data symbol will  affect the estimation accuracy of all the upcoming data symbols, with the strongest effect being on the adjacent data symbols.
To address this problem, in this section we introduce the  SSSgb$K$SE to reduce the error propagation effect of the proposed SSSSE, and hence, to improve its estimation accuracy. 

The basic idea of the proposed SSSgb$K$SE can be explained as follows. Upon the estimation of the current data symbol $a_k$, one can go back and improve the estimation accuracy of the previous $K$ data symbols $a_{k - 1}, \hdots, a_{k-K}$ based on the knowledge of the  current data  symbol $a_{k}$. Subsequently, we can re-estimate the current $k$th data symbol based on the improved estimation of the previous $K$ data symbols $a_{k - 1}, \hdots, a_{k-K}$. To explain the proposed SSSgb$K$SE in more details, let us rewrite the received $k$th sample value $y_k$ as
\begin{eqnarray}
y_{k} = \underbrace{G_{1, L} \: a_{k - L + 1} + \hdots + \underbrace{G_{1, K + 1} \: a_{k - K} + \hdots + G_{1, 2} \: a_{k - 1}}_{\textup{Previous $K$ symbols to be re-estimated}}}_{\textup{ISI from previous $L - 1$ symbols}} + \underbrace{G_{1,1} \: a_{k}}_{\textup{Current symbol to be re-estimated}} \nonumber \\ 
+ \underbrace{G_{1, 2} \: a_{k + 1} + \hdots + G_{1, L} \: a_{k + L - 1}}_{\textup{ISI from upcoming $L - 1$ symbols}}.
\end{eqnarray}
Hence, the improved re-estimation of the $(k - K)$th data symbol can be written as
\begin{eqnarray}
\hat{a}_{k - K} = {\rm{quantize}} \Big\{y_{k - K} \quad -  \underbrace{(G_{1,L} \: \hat{a}_{k - K - L + 1} + \hdots + G_{1,2} \: \hat{a}_{k - K - 1})}_{\textup{ISI from previous $L - 1$ data symbols of the $(k - K)$th data symbol}} \nonumber \\ - \underbrace{(G_{1, 2} \: \hat{\hat{a}}_{k - K + 1} + \hdots + G_{1, K + 1} \: \hat{\hat{a}}_{k})}_{\textup{ISI from upcoming $K$ data symbols of the $(k - K)$the data symbol}}\Big\}.
\end{eqnarray}
Similarly, the $k - 1, k - 2, \hdots, k - K + 1$ data symbols are re-estimated to improve their estimation accuracy. 
Accordingly, the $k$th data symbol can be re-estimated as
\begin{eqnarray}
\hat{a}_k = {\rm{quantize}} \Big\{y_k - \underbrace{(G_{1, L} \: \hat{a}_{k - L + 1} + \hdots + \underbrace{G_{1, K + 1} \: \hat{\hat{a}}_{k - K} +  \hdots + G_{1, 2} \: \hat{\hat{a}}_{k - 1}}_{\textup{ISI from previous $K$ data symbols with improved estimation accuracy}})}_{\textup{ISI from the previous $L - 1$ data symbols}}\Big\}.
\end{eqnarray}



\subsection{Proposed SSSgb$K$SE  and Complexity Analysis}
The proposed SSSgb$K$SE  is formally expressed as follows:

{\textit{Algorithm 3:} Proposed SSSgb$K$SE }
\begin{enumerate}
	\item \textbf{Input:} rRC pulses shape, its roll-off factor $\beta$, the time acceleration parameter $\tau$, and $K$.
	\item \textbf{If} $\beta$ and $\tau$ belong to the operation region shown in Table \ref{table:region} \textbf{then}
	\item \hspace{1cm} $\hat{a}_k = {\rm{quantize}} \left\{y_k - (G_{1, L} \: \hat{a}_{k - L + 1} + \hdots + G_{1, K + 1} \: \hat{\hat{a}}_{k - K} +  \hdots + G_{1, 2} \: \hat{\hat{a}}_{k - 1})\right\}$.
	\item \textbf{End}
\end{enumerate}

As discussed earlier, the key idea of the proposed SSSgb$K$SE is to re-estimate up to $K$ previous data symbols to improve the estimation accuracy of the current $k$th data symbol. The computational complexity of the proposed SSSgb$K$SE can be analyzed as follows:
\begin{itemize}
	\item To re-estimate the $(k-1)$th data symbol, we need $L-2$ additions and $L-1$ multiplications operations to remove the ISI from the previous $L-1$ data symbols; this is similar to the complexity of the proposed SSSSE. Additionally, $1$ multiplication operation is required to subtract the ISI from the upcoming $k$th data symbol. 
	\item To re-estimate the $(k - 2)$th data symbol, we need $L - 2$ additions and $L-1$ multiplications operations to remove the ISI from the previous $L-1$ data symbols in addition to 1 addition and 2 multiplication operations to subtract the ISI from the upcoming $(k - 1)$th and $k$th data symbols.
	\item The complexity of re-estimating up to previous $K$ data symbol can be done in the same manner. For instance, to re-estimate the $(k - K)$th data symbol we need   $L - 2$ additions and $L-1$ multiplications operations to remove the ISI from the previous $L-1$ data symbols in addition to $K - 1$ additions and $K$ multiplications operations. 
\end{itemize}
Hence, the required number of iterations for the proposed SSSgb$K$SE can be summarized as $K (L - 2) + 1 + 2 + \hdots + (K - 1)$ additions and $K (L - 1) +  1 + 2 + \hdots + K$ multiplications operations. The computational complexities of the proposed SSSSE and  SSSgb$K$SE algorithms are summarized in Table \ref{table:comp}.

\begin{table}[]
	\centering
	\caption{Computational complexity of the SSSSE and  SSSgb$K$SE algorithms.}
	\label{table:comp}
\begin{tabular}{r|c|c}
	\hline
	Algorithm & Number of addition operations & Number of multiplication operations \\ \hline
	SSSSE & $L - 2$ & $L - 1$ \\ \hline
	\multicolumn{1}{c|}{SSSgb$K$SE} & \multicolumn{1}{c|}{$K (L - 2) + \frac{K(K - 1)}{2}$} & \multicolumn{1}{c}{$K (L - 1) + \frac{K(K + 1)}{2}$} \\ \hline
\end{tabular}
\end{table}


The works in \cite{sugiura2013frequency, ishihara2016frequency} require a complexity of $\mathcal{O}(N)$ to calculate the MMSE coefficients of the FDE. This is in addition to a complexity of $\mathcal{O}(N \log(N))$ to perform the FFT and iFFT. Hence, the complexity of the FDEs in \cite{sugiura2013frequency, ishihara2016frequency} equals $\mathcal{O}(N) + \mathcal{O}(N \log(N))$ = $\mathcal{O}(N \log(N))$, i.e., the complexity is dominated by the FFT and iFFT operations. Such complexity is much higher than its counterparts of the proposed SSSSE and SSSgb$K$SE algorithms. 
\section{Simulation Results} \label{sec:results}

\begin{figure}[!t]
	\centering
	\includegraphics[width=1.00\textwidth]{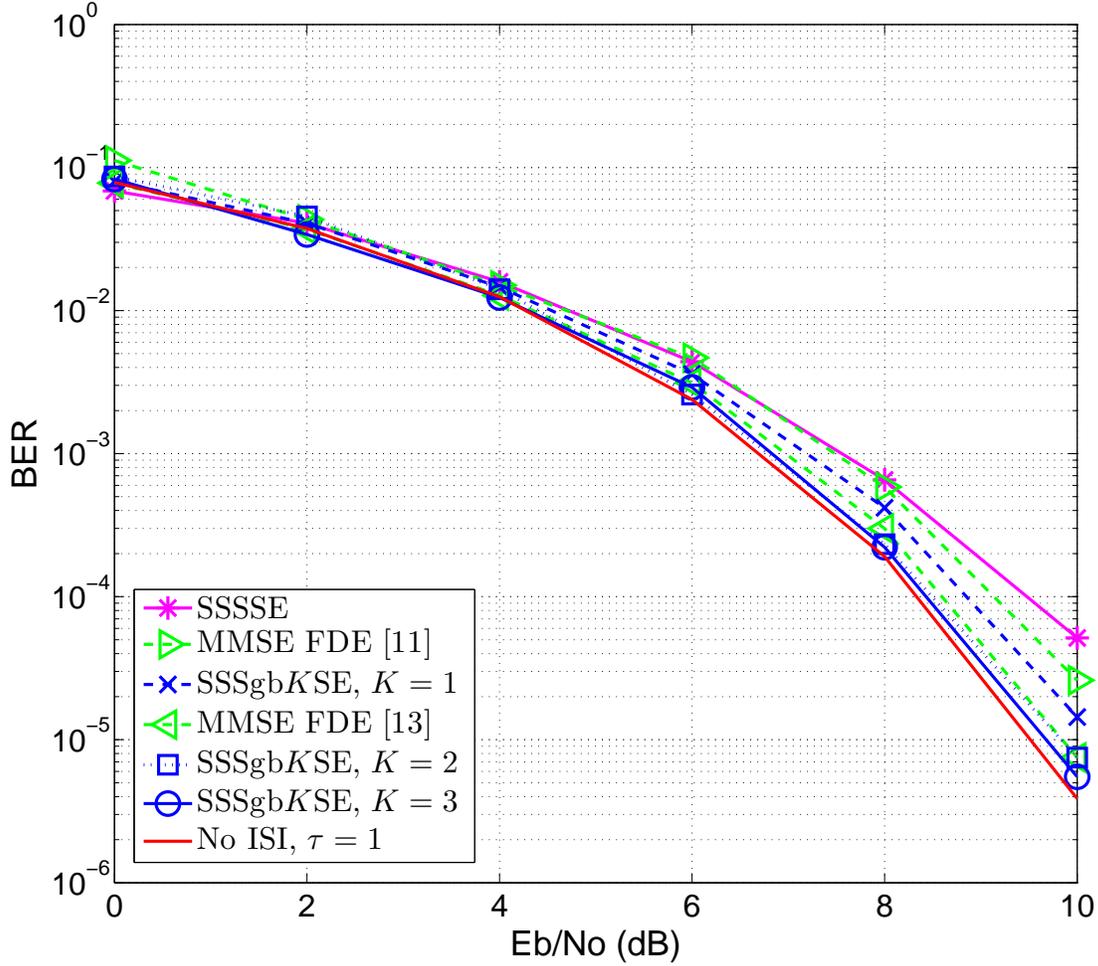}
	\caption{BER performance of {QPSK} FTN sequence estimation as a function of $\frac{E_b}{N_o}$ using the proposed SSSSE, proposed SSSgb$K$SE, and FDEs in \cite{sugiura2013frequency,ishihara2016frequency}  at $\beta = 0.3$ and SE  of $1.71$ bits/sec/Hz.}\label{fig:beta_03_tau_09}
\end{figure}

In this section, we evaluate the performance of the proposed SSSSE and SSSgb$K$SE in estimating transmit data symbols of FTN signaling. We employ a rRC filter with roll-off factors $\beta = 0.3$ and $0.5$ (unless otherwise mentioned), and we consider the data symbols to be drawn from the constellation of QPSK. We consider a transmission of data block of length $N = 1000$ data symbols and a cyclic prefix of length $\nu = 10$ symbols when simulating the works in \cite{sugiura2013frequency,ishihara2016frequency}. Hence, there is a loss of $1\%$  of the SE of the works in \cite{sugiura2013frequency,ishihara2016frequency} and it is considered in our simulations to have a fair comparison with the proposed SSSSE and SSSgb$K$SE schemes.
{ The SE is calculated as $\frac{\log_2 M}{(1 + \beta) \: \tau} \frac{N - \nu}{N}$, where $M$ is the constellation size.}

Fig. \ref{fig:beta_03_tau_09} depicts the BER of QPSK FTN signaling as a function of $\frac{E_b}{N_o}$ for the proposed SSSSE, SSSgb$K$SE for $K = 1, 2, 3$, and the FDEs in  \cite{sugiura2013frequency,ishihara2016frequency} for $\beta = 0.3$ and a SE of $1.71$ bits/sec/Hz. {This means that the value of $\tau$ used for our proposed SSSSE and SSSgb$K$SE is 0.9 and its value for the works in \cite{sugiura2013frequency,ishihara2016frequency} is set to 0.891.} As can be seen in Fig. \ref{fig:beta_03_tau_09}, increasing the value of $K$ improves the BER performance of the proposed SSSgb$K$SE as expected. Additionally, going back up to $K = 3$ data symbols approaches the optimal performance of the Nyquist ISI-free transmission for the given $\beta$ and SE values. Fig. \ref{fig:beta_03_tau_09} reveals that the proposed SSSgb$K$SE can achieve $\frac{1.71 - 1.54}{1.54}  = 11\%$ increase in the transmission rate without increasing the BER, the bandwidth, or the symbol energy, when compared to the Nyquist signaling (i.e., no ISI case). {Additionally, Fig. \ref{fig:beta_03_tau_09} shows the the proposed SSSgb$K$SE with $K = 1$ and $K = 2$ outperforms the works in  \cite{sugiura2013frequency} and \cite{ishihara2016frequency}, respectively.}

\begin{figure}[!t]
	\centering
	\includegraphics[width=1.00\textwidth]{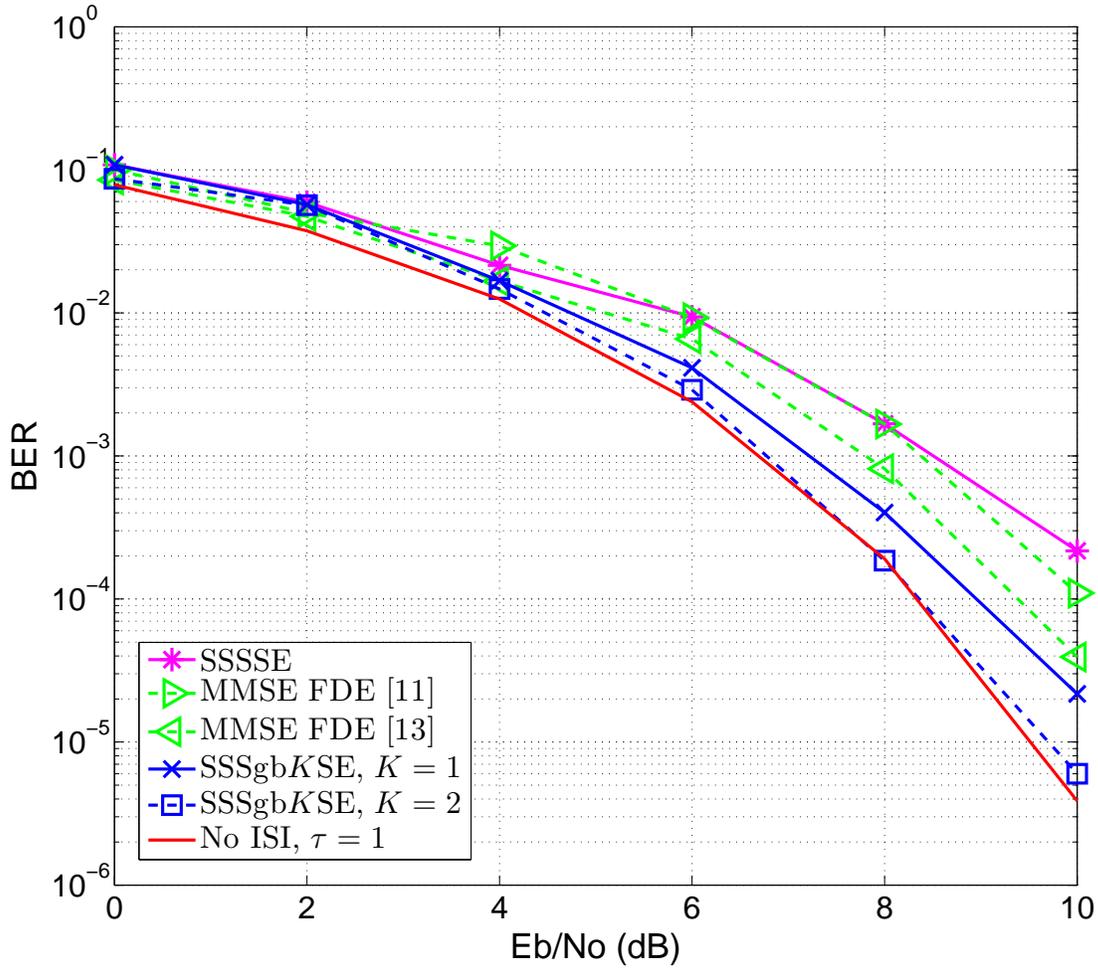}
	\caption{BER performance of {QPSK} FTN sequence estimation as a function of $\frac{E_b}{N_o}$ using the proposed SSSSE, proposed SSSgb$K$SE, and FDEs in \cite{sugiura2013frequency,ishihara2016frequency} at $\beta = 0.5$ and spectral efficiency bound of $1.67$ bits/sec/Hz.}\label{fig:beta_05_tau_08}
\end{figure}

Fig. \ref{fig:beta_05_tau_08} plots the BER of QPSK FTN as a function of $\frac{E_b}{N_o}$ for the proposed SSSSE, SSSgb$K$SE for $K = 1, 2$, and SDSE for $\beta = 0.5$ and a SE  of $1.67$ bits/sec/Hz. 
{This means that the value of $\tau$ used for our proposed SSSSE and SSSgb$K$SE is 0.8 and its value for the works in \cite{sugiura2013frequency,ishihara2016frequency} is set to 0.792.}
Similar to the previous discussion on Fig. \ref{fig:beta_03_tau_09}, going back for $K = 2$ data symbols at $\beta = 0.5$ and a SE  of $1.67$ bits/sec/Hz is enough to approach the performance of the Nyquist ISI-free transmission. One can infer from Fig. \ref{fig:beta_05_tau_08} that the proposed SSSgb$K$SE can achieve $25\%$ increase in the transmission rate without deteriorating the BER or increasing the bandwidth or the SNR, when compared to the Nyquist signaling. Additionally, the performance of the proposed SSSgb$K$SE with $K = 2$ surpasses the performance of the works in   \cite{sugiura2013frequency,ishihara2016frequency}.

\begin{figure}[!t]
	\centering
	\includegraphics[width=1.00\textwidth]{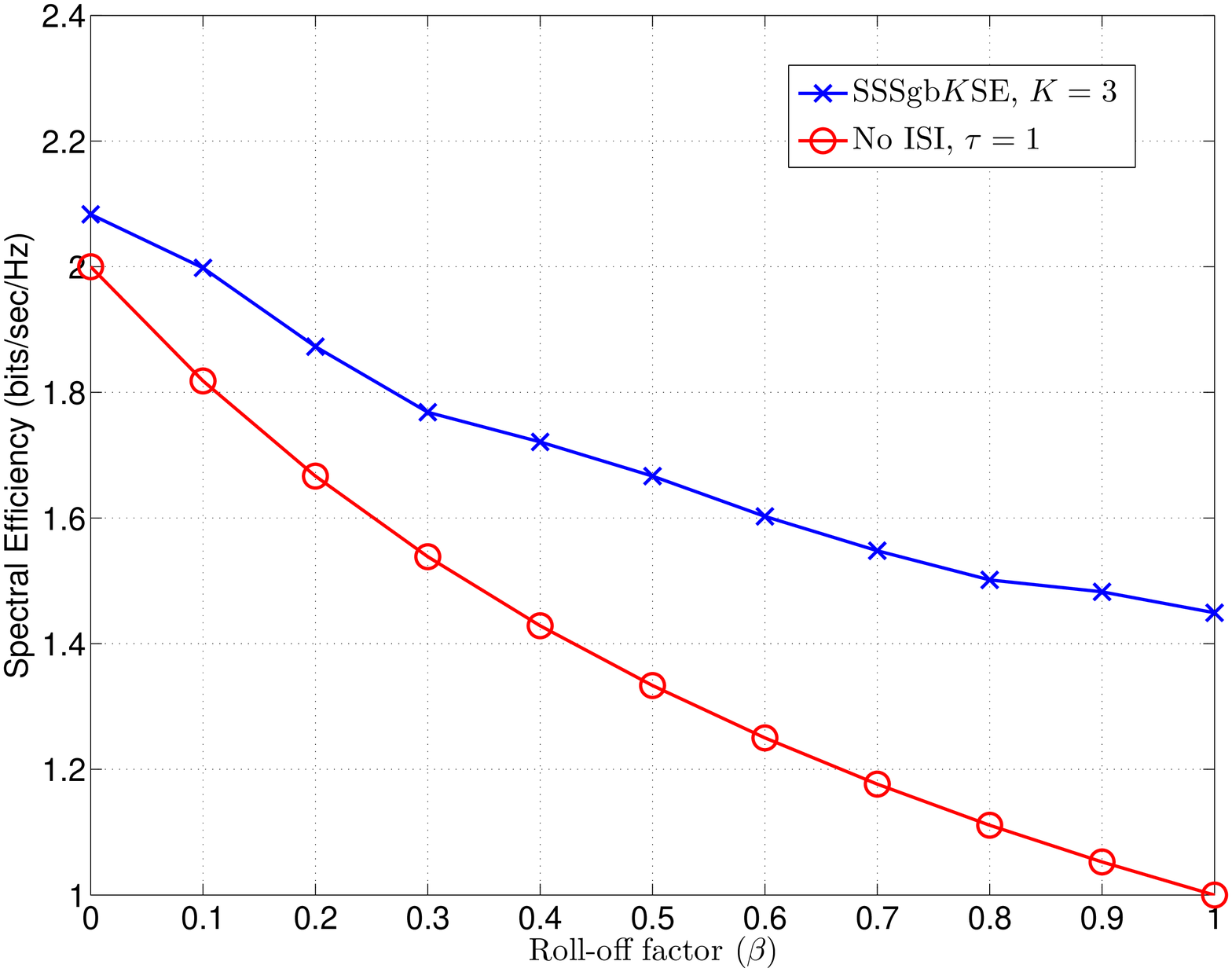}
	\caption{Spectral efficiency of QPSK Nyquist and FTN signaling as a function of $\beta$ using the proposed SSSgb$K$SE at BER = $10^{-4}$.}\label{fig:SE_2}
\end{figure}

Fig. \ref{fig:SE_2} plots the SE of QPSK Nyquist (i.e., no ISI and $\tau = 1$) and FTN signaling as a function of the roll-off factor $\beta$ at the same SNR and BER = $10^{-4}$. In order to have a fair comparison, the value of $\tau$ of the FTN signaling is selected to be the smallest value such that the proposed SSSgb$K$SE with $K = 3$ achieves the same BER = $10^{-4}$ of Nyquist signaling at the same SNR. As can be seen, the SE of FTN signaling is  higher than its counterpart of Nyquist signaling for all values of $\beta$. For instance, at $\beta = 0$ and $0.3$ the proposed SSSgb$K$SE  improves the SE by $4\%$ and $20.55\%$, respectively, for the same BER and SNR values, when compared to Nyquist signaling.
One can also infer from Fig.  \ref{fig:SE_2} that the FTN signaling exploits the excess bandwidth of the rRC pulse as the gain in SE of the proposed  SSSgb$K$SE  increases for increasing the value of $\beta$. 
Additionally, results revealed that the proposed SSSgb$K$SE can achieve SE higher than the maximum SE of Nyquist signaling (2 bit/s/Hz achieved at $\beta = 0$) for the range of $\beta \in [0, 0.1]$.

\section{Conclusion} \label{sec:conc}

FTN signaling is a promising non-orthogonal transmission technique capable of significantly increasing the spectral efficiency, when compared to the conventional Nyquist signaling. 
This paper presents the first attempt in the literature to detect FTN signaling on  a symbol-by-symbol basis, i.e., with very low computational complexity.
We  proposed two novel  sequence estimation techniques, namely, SSSSE and SSSgb$K$SE, to estimate the transmit data symbols of BPSK and QPSK FTN signaling. 
In particular, based on concepts of successive interference cancellation we found the  operating region boundary (defined by the rRC pulse shape, its roll-off factor, and the time acceleration parameter), where the proposed SSSSE and SSSgb$K$SE can perfectly estimate the transmit data symbols for noise-free transmission.

For noisy transmission, the proposed SSSSE estimates the data symbols on  a symbol-by-symbol basis, with a significant reduction in the  computational complexity when compared to all the sequence estimation techniques from the literature. 
To overcome the error propagation effect in the SSSSE, the proposed SSSgb$K$SE can go-back to re-estimate up to $K$ data symbols, based on the knowledge of the current data symbol, and accordingly improves the estimation accuracy of the current data symbol based on the re-estimation of the previous $K$ data symbols. The proposed schemes are of low complexity. More specifically, the proposed SSSSE requires additional $L - 2$ additions and $L-1$ multiplications  operations when compared to Nyquist signaling; while the proposed SSSgb$K$SE requires additional $K (L - 2) + \frac{K(K - 1)}{2}$ additions and $K (L - 1) + \frac{K(K + 1)}{2}$ multiplications operations.
Simulation results showed that  the proposed techniques are suitable for low ISI scenarios and can achieve up to $11.11\%$ and $25\%$ increase in the transmission rate at $\beta = 0.3$ and $0.5$, respectively, and up to $4\%$ and $20.55\%$ improvement in the spectral efficiency at $\beta = 0$ and $0.3$, respectively, when compared to Nyquist signaling, for the same SNR and BER. 
We showed that for low ISI scenarios choosing  $K = 2$ or $3$ data symbols is sufficient to improve the BER performance. Additionally, results revealed that the proposed SSSgb$K$SE can achieve spectral efficiency higher than the maximum spectral efficiency of QPSK Nyquist signaling (2 bit/s/Hz achieved at $\beta = 0$) for the range of $\beta \in [0, 0.1]$.

%
\IEEEpeerreviewmaketitle




%



\ifCLASSOPTIONcaptionsoff
  \newpage
\fi



%

\bibliographystyle{IEEEtran}
\bibliography{IEEEabrv,mybib_file}

\begin{thebibliography}{10}
\providecommand{\url}[1]{#1}
\csname url@samestyle\endcsname
\providecommand{\newblock}{\relax}
\providecommand{\bibinfo}[2]{#2}
\providecommand{\BIBentrySTDinterwordspacing}{\spaceskip=0pt\relax}
\providecommand{\BIBentryALTinterwordstretchfactor}{4}
\providecommand{\BIBentryALTinterwordspacing}{\spaceskip=\fontdimen2\font plus
\BIBentryALTinterwordstretchfactor\fontdimen3\font minus
  \fontdimen4\font\relax}
\providecommand{\BIBforeignlanguage}[2]{{%
\expandafter\ifx\csname l@#1\endcsname\relax
\typeout{** WARNING: IEEEtran.bst: No hyphenation pattern has been}%
\typeout{** loaded for the language `#1'. Using the pattern for}%
\typeout{** the default language instead.}%
\else
\language=\csname l@#1\endcsname
\fi
#2}}
\providecommand{\BIBdecl}{\relax}
\BIBdecl

\bibitem{saltzberg1968intersymbol}
B.~Saltzberg, ``Intersymbol interference error bounds with application to ideal
  bandlimited signaling,'' \emph{{IEEE} Trans. Inf. Theory}, vol.~14, no.~4,
  pp. 563--568, Jul. 1968.

\bibitem{lucky1970decision}
R.~Lucky, ``{Decision feedback and faster-than-Nyquist transmission},'' in
  \emph{Proc. IEEE International Symposium on Information Theory (ISIT)}, Jun.
  1970, pp. 15--19.

\bibitem{salz1973optimum}
J.~Salz, ``Optimum mean-square decision feedback equalization,'' \emph{Bell
  Syst. Tech. J.}, vol.~52, no.~8, pp. 1341--1373, Oct. 1973.

\bibitem{mazo1975faster}
J.~Mazo, ``{Faster-than-Nyquist signaling},'' \emph{Bell Syst. Tech. J.},
  vol.~54, no.~8, pp. 1451--1462, Oct. 1975.

\bibitem{foschini1984contrasting}
G.~J. Foschini, ``{Contrasting performance of faster binary signaling with
  QAM},'' \emph{Bell Syst. Tech. J.}, vol.~63, no.~8, pp. 1419--1445, Oct.
  1984.

\bibitem{rusek2006cth04}
F.~Rusek and J.~B. Anderson, ``{On information rates for faster than Nyquist
  signaling},'' in \emph{Proc. IEEE Global Communication Conference
  (GLOBECOM)}, Dec. 2006, pp. 1--5.

\bibitem{rusek2009constrained}
------, ``{Constrained capacities for faster-than-Nyquist signaling},''
  \emph{{IEEE} Trans. Inf. Theory}, vol.~55, no.~2, pp. 764--775, Feb. 2009.

\bibitem{liveris2003exploiting}
A.~D. Liveris and C.~N. Georghiades, ``{Exploiting faster-than-Nyquist
  signaling},'' \emph{{IEEE} Trans. Commun.}, vol.~51, no.~9, pp. 1502--1511,
  Sep. 2003.

\bibitem{prlja2008receivers}
A.~Prlja, J.~B. Anderson, and F.~Rusek, ``{Receivers for faster-than-Nyquist
  signaling with and without turbo equalization},'' in \emph{Proc. IEEE
  International Symposium on Information Theory}, Jul. 2008, pp. 464--468.

\bibitem{anderson2009new}
J.~B. Anderson, A.~Prlja, and F.~Rusek, ``New reduced state space {BCJR}
  algorithms for the {ISI} channel,'' in \emph{Proc. IEEE International
  Symposium on Information Theory (ISIT)}, Jun. 2009, pp. 889--893.

\bibitem{sugiura2013frequency}
S.~Sugiura, ``{Frequency-domain equalization of faster-than-Nyquist
  signaling},'' \emph{IEEE Wireless Commun. Lett.}, vol.~2, no.~5, pp.
  555--558, Oct. 2013.

\bibitem{sugiura2015frequency}
S.~Sugiura and L.~Hanzo, ``{Frequency-domain-equalization-aided iterative
  detection of faster-than-Nyquist signaling},'' \emph{{IEEE} Trans. Veh.
  Technol.}, vol.~64, no.~5, pp. 2122--2128, May 2015.

\bibitem{ishihara2016frequency}
T.~Ishihara and S.~Sugiura, ``{Frequency-domain equalization aided iterative
  detection of faster-than-Nyquist signaling with noise whitening},'' in
  \emph{Proc. IEEE International Conference on Communications (ICC)}, May 2016,
  pp. 1--6.

\bibitem{dinis2015hybrid}
R.~Dinis, B.~Cunha, F.~Ganhao, L.~Bernardo, R.~Oliveira, and P.~Pinto, ``{A
  Hybrid ARQ Scheme for Faster than Nyquist Signaling with Iterative
  Frequency-Domain Detection},'' in \emph{Proc. IEEE Vehicular Technology
  Conference (VTC Spring)}, 2015, pp. 1--5.

\bibitem{lucciardi2016trade}
J.-A. Lucciardi, N.~Thomas, M.-L. Boucheret, C.~Poulliat, and G.~Mesnager,
  ``{Trade-off between spectral efficiency increase and PAPR reduction when
  using FTN signaling: Impact of non linearities},'' in \emph{Proc. IEEE
  International Conference on Communications (ICC)}, May 2016, pp. 1--7.

\bibitem{le2015faster}
C.~Le, M.~Fuhrwerk, M.~Schellmann, and J.~Peissig, ``Faster than nyquist--an
  enabler for achieving maximum spectral efficiency in coexistence scenarios?''
  in \emph{Proc. European Signal Processing Conference (EUSIPCO)}, Aug. 2015,
  pp. 2142--2146.

\end{thebibliography}

%




\end{document}